\documentclass{tlp}
\usepackage{ifthen}
\usepackage{url}
\usepackage{swipl}

\submitted{28 April 2006}
\revised{23 August 2007}
\accepted{18 October 2007}

\usepackage[pdftex]{graphicx}
\DeclareGraphicsExtensions{.pdf,.jpg,.png}
\graphicspath{{figs/}{./}}
\newcommand{\tag}[1]{\texttt{#1}}
\newcommand{\fnurl}[1]{\footnote{\url{#1}}}
\begin{document}
\bibliographystyle{acmtrans}

\long\def\comment#1{}

\title{SWI-Prolog and the Web}

\author[J. Wielemaker, Z. Huang and L. van der Meij]
{JAN WIELEMAKER \\
Human-Computer Studies laboratory \\
University of Amsterdam\\
Matrix I \\
Kruislaan 419 \\
1098 VA, Amsterdam \\
The Netherlands \\
\email{wielemak@science.uva.nl}
\and
ZHISHENG HUANG, LOURENS VAN DER MEIJ \\
Computer Science Department\\
Vrije University Amsterdam\\
De Boelelaan 1081a \\
1081 HV, Amsterdam\\
The Netherlands\\
\email{huang,lourens@cs.vu.nl}
}

\pagerange{\pageref{firstpage}--\pageref{lastpage}}
\volume{\textbf{?} (?):}
\jdate{August 2007}
\setcounter{page}{1}
\pubyear{2007}

\maketitle

\label{firstpage}

\begin{abstract}
Prolog is an excellent tool for representing and manipulating data
written in formal languages as well as natural language.  Its safe
semantics and automatic memory management make it a prime candidate for
programming robust Web services.  

Where Prolog is commonly seen as a component in a Web application that
is either embedded or communicates using a proprietary protocol, we
propose an architecture where Prolog communicates to other components in
a Web application using the standard HTTP protocol. By avoiding
embedding in external Web servers development and deployment become much
easier. To support this architecture, in addition to the transfer
protocol, we must also support parsing, representing and generating the
key Web document types such as HTML, XML and RDF.

This paper motivates the design decisions in the libraries and
extensions to Prolog for handling Web documents and protocols. The
design has been guided by the requirement to handle large documents
efficiently. The described libraries support a wide range of Web
applications ranging from HTML and XML documents to Semantic Web RDF
processing.

The benefits of using Prolog for Web related tasks is illustrated using
three case studies.
\end{abstract}



\section{Introduction}


The Web is an exciting place offering new opportunities to artificial
intelligence, natural language processing and Logic Programming.
Information extraction from the Web, reasoning in Web applications and
the Semantic Web are just a few examples. We have deployed Prolog in Web
related tasks over a long period. As most of the development on
SWI-Prolog takes place in the context of projects that require new
features, the system and its libraries provide extensive support for Web
programming.

There are two views on deploying Prolog for Web related tasks. In the
most commonly used view, Prolog acts as an embedded component in a
general Web processing environment. In this role it generally provides
reasoning tasks such as searching or configuration within constraints.
Alternatively, Prolog itself can act as a stand-alone HTTP server as also
proposed by ECLiPSe \cite{Eclipse-http}. In this view it is a component
that can be part of any of the layers of the popular three-tier
architecture for Web applications. Components generally exchange
XML if used as part of the backend or middleware services and HTML if
used in the presentation layer.

The latter view is in our vision more attractive. Using HTTP and XML
over HTTP, the service is cleanly isolated using standard protocols
rather than proprietary communication. Running as a stand-alone
application, the attractive interactive development nature of Prolog can
be maintained much more easily than embedded in a C, C++, Java or C\#
application. Using HTTP, automatic testing of the Prolog components can
be done using any Web oriented test framework. HTTP allows Prolog to be
deployed in any part of the service architecture, including the
realisation of complete Web applications in one or more Prolog
processes.

When deploying Prolog in a Web application using HTTP, we must not only
implement the HTTP transfer protocol, but also support parsing,
representing and generating the important document types used on the
Web, especially HTML, XML and RDF. Note that, being widely used open
standards, supporting these document types is also valuable outside the
context of Web applications.

This paper gives an overview of the Web infrastructure we have
realised. Given the range of libraries and Prolog extensions that
facilitate Web applications we cannot describe them in detail. Details
on the library interfaces can be found in the manuals available from the
SWI-Prolog Web site.\fnurl{http://www.swi-prolog.org} Details on the
implementation are available in the source distribution. The aim of this
paper is to give an overview of the required infrastructure to use
Prolog for realizing Web applications where we concentrate on
scalability and performance. We describe our decisions for representing
Web documents in Prolog and outline the interfaces provided by our
libraries.

The benefits of using Prolog for Web related tasks are illustrated using
three case studies: 1) SeRQL, an RDF query language for meta data
management, retrieval and reasoning; 2) XDIG, an eXtended Description
Logic interface, which provides ontology management and reasoning by
processing DIG XML documents and communicating to external DL reasoners;
and 3) A faceted browser on Semantic Web databases integrating meta-data
from multiple collections of art-works. This case study serves as a
complete Semantic Web application serving the end-user.

This paper is organized as follows. \Secref{markupdom} to
\secref{rdfdoc} describe reading, writing and representation of
Web related documents. \Secref{http} describes our HTTP client and
server libraries. \Secref{prologext} describes extensions to the Prolog
language that facilitate use in Web applications. \Secref{serql} to
\secref{facetb} describe the case studies.


\section{Parsing and representing XML and HTML documents}
\label{sec:markupdom}


The core of the Web is formed by document standards and exchange
protocols. Here we describe tree-structured documents transferred as
SGML or XML. HTML, an SGML application, is the most commonly used
document format on the Web. HTML represents documents as a tree using a
fixed set of \jargon{elements} (tags), where the SGML DTD (Document Type
Declaration) puts constraints on how elements can be nested. Each node
in the hierarchy has a name (the element-name), a set of name-value
pairs known as its attributes and \jargon{content}, a sequence of
sub-elements and text (data).

XML is a rationalisation of SGML using the same tree-model, but removing
many rarely used features as well as abbreviations that were introduced
in SGML to make the markup easier to type and read by humans. XML
documents are used to represent text using custom application-oriented
tags as well as a serialization format for arbitrary data exchange
between computers. XHTML is HTML based on XML rather than SGML.

The first SGML parser for SWI-Prolog was created by Anjo Anjewierden
based on the SP parser\footnote{\url{http://www.jclark.com/sp/}}. A
stable Prolog term-representation for SGML/XML trees plays a similar
role as the DOM (\jargon{Document Object Model}) representation in
use in the object-oriented world. The term-structure we use is described
in \figref{dom}. Some issues have been subject to debate.

\begin{figure}
\figrule
\begin{tabular}{lrl}
\bnfmeta{document}	\isa list-of \bnfmeta{content} \\
\bnfmeta{content}	\isa \bnfmeta{element} $\mid$ \bnfmeta{pi} $\mid$ \bnfmeta{cdata} $\mid$ \bnfmeta{sdata} $\mid$ \bnfmeta{ndata} \\
\bnfmeta{element}	\isa element(\bnfmeta{tag}, list-of \bnfmeta{attribute}, list-of \bnfmeta{content}) \\
\bnfmeta{attribute}	\isa \bnfmeta{name} = \bnfmeta{value} \\
\bnfmeta{pi}		\isa pi(\bnfmeta{atom})  \\
\bnfmeta{sdata}		\isa sdata(\bnfmeta{atom})  \\
\bnfmeta{ndata}		\isa ndata(\bnfmeta{atom}) \\
\bnfmeta{cdata}, \bnfmeta{name}	\isa \bnfmeta{atom} \\
\bnfmeta{value}		\isa \bnfmeta{svalue} $\mid$ list-of \bnfmeta{svalue} \\
\bnfmeta{svalue}	\isa \bnfmeta{atom} $\mid$ \bnfmeta{number}
\end{tabular}
    \caption{SGML/XML tree representation in Prolog.   The notation
	     list-of \bnfmeta{x} describes a Prolog list of terms of type \bnfmeta{x}.}
    \label{fig:dom}
\figrule
\end{figure}

\begin{itemize}
    \item Representation of text by a Prolog atom is biased by the use of
          SWI-Prolog which has no length-limit on atoms and atoms that
	  can represent Unicode text as motivated in
	  \secref{unicode}. At the same time SWI-Prolog stacks are
	  limited to 128MB each. Using atoms only the structure of the
	  tree is represented on the stack, while the bulk of the data
	  is stored on the unlimited heap. Using lists of character
	  codes is another possibility adopted by both PiLLoW
	  \cite{PiLLoW} and ECLiPSe \cite{Eclipse-http}. Two
	  observations make lists less attractive: lists use two cells
	  per character while practical experience shows text is
	  frequently processed as a unit only. For (HTML) text-documents
	  we profit from the compact representation of atoms. For XML
	  documents representing serialized data-structures we profit
	  from frequent repetition of the same value.

    \item Attribute values of multi-value attributes (e.g.\ 
 	  \tag{NAMES}) are returned as a Prolog list.   This implies
	  the DTD must be available to get unambiguous results.  With
	  SGML this is always true, but not with XML.

    \item Optionally attribute values of type \tag{NUMBER} or \tag{NUMBERS}
	  are mapped to Prolog numbers.  In addition to the DTD issues
	  mentioned above, this conversion also suffers from possible
	  loss of information. Leading zeros and different floating
	  point number notations used are lost after conversion. Prolog
	  systems with bounded arithmetic may also not be able to
	  represent all values. Still, automatic conversion is useful in
	  many applications, especially those involving serialized
	  data-structures.

    \item Attribute values are represented as \arg{Name}=\arg{Value}.  Using
    	  \mbox{\arg{Name}({Value})} is an alternative. The
	  \arg{Name}=\arg{Value} representation was chosen for its
	  similarity to the SGML notation and because it avoids the
	  need for univ (\verb|=..|) for processing argument-lists.
\end{itemize}

\paragraph{Implementation}

The SWI-Prolog SGML/XML parser is implemented as a C-library that has
been built from scratch to create a lightweight parser. Total
source is 11,835 lines. The parser provides two interfaces. Most natural
to Prolog is \term{load_structure}{+Src, -DOM, +Options} which parses a
Prolog stream into a term as described above. Alternatively,
\index{sgml_parse/2}\predref{sgml_parse}{2} provides an \jargon{event-based} parser making call-backs
on Prolog for the SGML events. The call-back mode can deal with
unbounded documents in streaming mode. It can be mixed with the
term-creation mode, where the handler for \emph{begin} calls the parser
to create a term-representation for the content of the element. This
feature is used to process long files with a repetitive record structure
in limited memory. \Secref{rdfio} describes how this is used to process
RDF documents.

Full documentation is available from
\url{http://www.swi-prolog.org/packages/sgml2pl.html}
The SWI-Prolog SGML parser has been adopted by XSB Prolog.

\section{Generating Web documents}
\label{sec:genml}

There are many approaches to generating Web pages from programs in
general and Prolog in particular. We believe the preferred choice
depends on various aspects.

\begin{itemize}
    \item
How much of the document is generated from dynamic data and how much is
static?  Pages that are static except for a few strings are best
generated from a template using variable substitution. Pages
consisting of a table generated from dynamic data are best entirely
generated from the program.

    \item
For program generated pages we can choose between direct printing and
generating using a language-native syntax, for example
\verb|format('<b|\verb|>bold</b>')| or \verb|print_html(b(bold))|. The
second approach can guarantee well-formed output, but the first requires
the programmer to learn about \index{format/3}\predref{format}{3} only.

    \item
Documents that contain a significant static part are best represented in
the markup language where special constructs insert program-generated
parts. A popular approach implemented by PHP\fnurl{www.php.net} and
ASP\fnurl{www.microsoft.com} is to add a reserved element such as
\bnfmeta{script} and use the SGML/XML \jargon{programming instruction} written
as \verb"<?...?>". The obvious name PSP (Prolog Server Pages) is in use
by various projects taking this approach.%
 \footnote{\url{http://www.prologonlinereference.org/psp.psp},\\
	   \url{http://www.benjaminjohnston.com.au/template.prolog?t=psp},\\
	   \url{http://www.ifcomputer.com/inap/inap2001/program/inap_bartenstein.ps}}
Another approach is
PWP\fnurl{http://www.cs.otago.ac.nz/staffpriv/ok/pwp.pl} (Prolog
Well-formed Pages).  It is based on the principle that the source
is well-formed XML and interacts with Prolog through additional
attributes.  Output is guaranteed to be well-formed XML.
Our infrastructure does not yet include any of these approaches.

    \item
Page \jargon{transformation} is realised by parsing the original
document into its tree representation, managing the tree and writing a
new document from the tree. Managing the source-text directly is not
reliable as due to character encoding choice, entity usage and SGML
abbreviations there are many different source-texts that represent the
same tree. The \index{load_structure/3}\predref{load_structure}{3} predicate described in
\secref{markupdom} together with output primitives from the library
\file{sgml_write.pl} provide this functionality.  The XDIG case study
described in \secref{xdig} follows this approach.
\end{itemize}

\subsection{Generating documents using DCG}
\label{sec:htmlwrite}

The traditional method for creating Web documents is using print
routines such as \index{write/1}\predref{write}{1} or \index{format/2}\predref{format}{2}. Although simple and easily
explained to novices, the approach has serious drawbacks from a software
engineering point of view. In particular the user is responsible for
HTML quoting, character encoding issues and proper nesting of HTML
elements. Automated validation is virtually impossible using this
approach.

Alternatively we can produce a DOM term as described in
\secref{markupdom} and use the library \file{sgml_write.pl} to create
the HTML or XML document. Such documents are guaranteed to use proper
nesting of elements, escape sequences and character encoding. The terms
however are big, deeply nested and hard to read and write. Prolog allows
them to be built from skeletons containing variables. This approach is
taken by PiLLoW (\secref{pillow}) to control the complexity. In our
opinion, the result is not optimal due to the unnatural order of
statements as illustrated in \figref{pillowwrite}. PiLLoW has partly
overcome this shortcoming by defining a large number of `utility terms'
that are translated in a special way, as discussed in section 6.2 of
\cite{PiLLoW}.

\begin{figure}
\figrule
\begin{verbatim}
        ...,
        mkthumbnail(URL, Caption, ThumbNail),
        output_html([ h1("Photo gallery"),
                      ThumbNail
                    ]).

mkthumbnail(URL, Caption, Term) :-
        Term = table([ tr(td([halign=center], img([src=URL],[]))),
                       tr(td([halign=center], Caption))
                     ])
\end{verbatim}

\noindent
    \caption{Building PiLLoW terms}
    \label{fig:pillowwrite}
\figrule
\end{figure}

We introduced a DCG rule html//1.%
    \footnote{The notation \bnfmeta{name}//\bnfmeta{arity} refers to the grammar rule
	      \bnfmeta{name} with the given \bnfmeta{arity}, and consequently the
	      predicate \bnfmeta{name} with arity \bnfmeta{arity}$+2$.}
This rule translates proper trees into a list of high-level HTML/XML
commands that are handed to \index{html_print/1}\predref{html_print}{1} to realise proper quoting,
character encoding and layout. The intermediate format is of no concern
to the user and similar in structure to the PiLLoW representation
without using environments. Generated from the tree representation
however, consistent opening and closing of elements is guaranteed. In
addition to variable substitution which is provided by Prolog we allow
calling rules. Rules are invoked by a term \verb|\|\arg{Rule} embedded
in the argument of html//1. \Figref{affiliation} illustrates our
approach. Note that any reusable part of the page generation can easily
be translated into a DCG rule and the difference between direct
translation of terms to HTML and rule-invocation is eminent.

\begin{figure}
\figrule
\begin{verbatim}
affiliation_table :-
        findall(Name-Aff, affiliation(Name, Aff), Pairs0),
        keysort(Pairs0, Pairs),
        reply_page(table([border(2),align(center)],
                         [ tr([th('Name'), th('Affiliation')])
                         | \affiliations(Pairs)
                         ])).

affiliations([]) --> [].
affiliations([H|T]) --> affiliation(H), affiliations(T).

affiliation(Name-Aff) --> html(tr(td(Name), td(Aff))).

% database
affiliation(wielemaker, uva).
affiliation(huang, vu).
affiliation('van der meij', vu).

% Page template
reply_page(Term) :-
        format('Content-type: text/html~n~n'),
        phrase(html(Term), Tokens),
        print_html(Tokens).
\end{verbatim}

\noindent
    \caption{Library html_write.pl in action}
    \label{fig:affiliation}
\figrule
\end{figure}

In our current implementation rules are called using meta-calling from
html//1. Using \index{term_expansion/2}\predref{term_expansion}{2} it is straightforward to move the rule
invocation out of the term, using variable substitution similar to
PiLLoW. It is also possible to recursively expand the generated tree
and validate it to the HTML DTD at compile-time and even insert omitted
tags at compile-time to generate valid XHMTL from an incomplete
specification.  An overview of the argument to html//1 is
given in \figref{html1}.

\begin{figure}
\figrule
\begin{tabular}{lrl}
\bnfmeta{html}		\isa list-of \bnfmeta{content} $\mid$ \bnfmeta{content} \\
\bnfmeta{content}	\isa \bnfmeta{atom} \\
		\ora \bnfmeta{tag}(list-of \bnfmeta{attribute}, \bnfmeta{html}) \\
		\ora \bnfmeta{tag}(\bnfmeta{html}) \\
		\ora \verb|\|\bnfmeta{rule} \\
\bnfmeta{attribute}	\isa \bnfmeta{name}(\bnfmeta{value}) \\
\bnfmeta{tag}, \bnfmeta{entity}	\isa \bnfmeta{atom} \\
\bnfmeta{value}		\isa \bnfmeta{atom} $\mid$ \bnfmeta{number} \\
\bnfmeta{rule}		\isa \bnfmeta{callable}
\end{tabular}
    \caption{The html//1 argument specification}
    \label{fig:html1}
\figrule
\end{figure}

\subsection{Comparison with PiLLoW}
\label{sec:pillow}

The PiLLoW library \cite{PiLLoW} is a well established framework
for Web programming based on Prolog. PiLLoW defines \index{html2terms/2}\predref{html2terms}{2},
converting between an HTML string and a document represented as a
Herbrand term. There are fundamental differences between PiLLoW and the
primitives described here.

\begin{itemize}
    \item
PiLLoW creates an HTML document from a Herbrand term that is passed
to \index{html2terms/2}\predref{html2terms}{2}. Complex terms are composed of partial terms passed as
Prolog variables. Frequent HTML constructs are supported using reserved
terms using dedicated processing. We use DCGs and the \verb|\|\arg{Rule}
construct, which makes it eminent which terms directly refer to HTML
elements and which function as a `macro'. In addition, the user can
define application-specific reusable fragments in a uniform way.

    \item
The PiLLoW parser does not create the SGML document tree. It does not
insert omitted tags, default attributes, etcetera. As a result, HTML
documents that differ only in omitted tags and whether or not default
attributes are included in the source produce different terms. In our
approach the term representation is equivalent, regardless of the input
document. This is illustrated in \figref{tabledom}. Having a canonical
DOM representation greatly simplifies processing parsed HTML documents.
\end{itemize}

\begin{figure}
\figrule
\begin{verbatim}
[env(table, [], [tr$[], td$[], "Hello"])]
\end{verbatim}

\noindent
\begin{verbatim}
[element(table, [],
         [ element(tbody, [],
                   [ element(tr, [],
                            [ element(td, [rowspan='1', colspan='1'],
                                      ['Hello'])])])])]
\end{verbatim}

\noindent
    \caption{Term representations for
	     \texttt{\string<table\mbox{}><tr\mbox{}><td\mbox{}>Hello</table>} in
	     PiLLoW (top) and our parser (bottom).  Our parser completes
	     the \tag{tr} and \tag{td} environments, inserts the omitted
	     \tag{tbody} element and inserts the defaults for the
	     \tag{rowspan} and \tag{colspan} attributes}
    \label{fig:tabledom}
\figrule
\end{figure}

\section{RDF documents}
\label{sec:rdfdoc}

\begin{figure}
    \centerline{\includegraphics[width=0.7\linewidth]{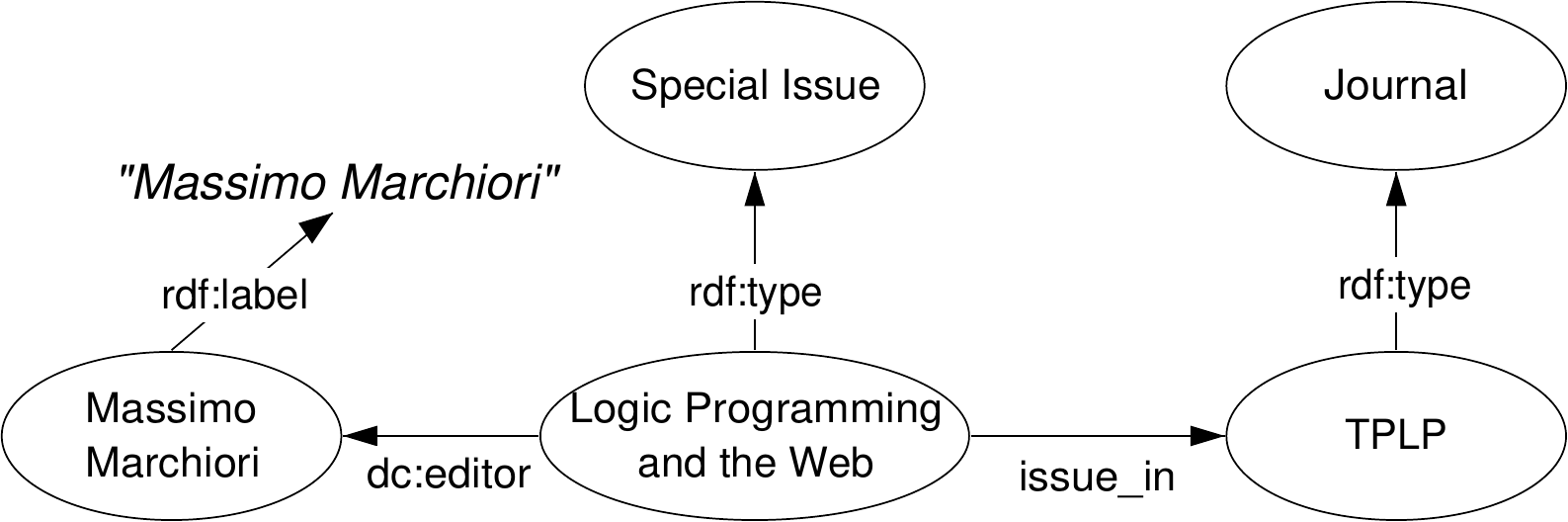}}
    \caption{Sample RDF graph.  Ellipses are vertices representing
	     URIs.  Quoted text is a literal.  Edges are labelled with
	     URIs.}
    \label{fig:triple}
\end{figure}

Where the datamodel of both HTML and XML is a tree-structure with
attributes, the datamodel of the Semantic Web (SW)
RDF\fnurl{http://www.w3.org/RDF/} language consists of
\mbox{\{\arg{Subject}, \arg{Predicate}, \arg{Object}\}}
\jargon{triples}. Both \arg{Subject} and \arg{Predicate} are
\jargon{URI}s.%
    \footnote{URI: \jargon{Uniform Resource Identifier} is
	      like a URL, but need not refer to an existing
	      resource on the Web.}
\arg{Object} is either a URI or a \jargon{Literal}. As the \arg{Object}
of one triple can be the \arg{Subject} of another, a set of triples
forms a graph, where each edge is labelled with a URI (the
\arg{Predicate}) and each vertex is either a URI or a literal. Literals
have no out-going edges. \Figref{triple} illustrates this.

A number of languages are layered on top of the RDF triple model. RDFS
provides a frame-based representation. The
OWL-dialects\fnurl{http://www.w3.org/2004/OWL/} provide three
increasingly complex Web ontology languages.
SWRL\fnurl{http://www.w3.org/Submission/SWRL/} is a proposal for a rule
language. The W3C standard for exchanging these triple models is an XML
application known as RDF/XML.

As there are multiple XML tree representations for the same triple-set,
RDF documents cannot be processed at the level of the XML-DOM as
described in \secref{markupdom}. A triple- or graph-based structure is
the most natural choice for representating an RDF document in Prolog.
First we must decide on the representation of URIs and literals. As a
URI is a string and the only operation defined on URIs by SW languages
is equivalence test, using a Prolog atom is an obvious choice. One may
consider using a term \mbox{\bnfmeta{namespace}:\bnfmeta{localname}}, but given that
decomposing a URI into its namespace and localname is only relevant
during I/O we consider this an inferior choice. The RDF library comes
with a compile-time rewrite mechanism based on \index{goal_expansion/2}\predref{goal_expansion}{2} that
allows for writing resources in Prolog sourcetext as
\mbox{\bnfmeta{ns}:\bnfmeta{local}}. Literals are expressed as \term{literal}{Value}.
The full type description is in \figref{rdftypes}.

\begin{figure}
\figrule
\begin{tabular}{lrl}
\bnfmeta{subject}, \bnfmeta{predicate}	\isa \bnfmeta{URI} \\
\bnfmeta{object}	\isa \bnfmeta{URI} \\
		\ora literal(\bnfmeta{lit_value}) \\
\bnfmeta{lit_value}	\isa \bnfmeta{text} \\
		\ora lang(\bnfmeta{langid}, \bnfmeta{text}) \\
		\ora type(\bnfmeta{URI}, \bnfmeta{text}) \\
\bnfmeta{URI}, \bnfmeta{text}	\isa \bnfmeta{atom} \\
\bnfmeta{langid}	\isa \bnfmeta{atom} (ISO639)
\end{tabular}
    \caption{RDF types in Prolog.}
    \label{fig:rdftypes}
\figrule
\end{figure}

The typical SW use-scenario is to `harvest' triples from multiple
sources and collect them in a database before reasoning with them.
Prolog can represent data as a Herbrand term on the stack or as
predicates in the database. Given the relatively static nature of the
RDF data as well as desired access from multiple threads, using the Prolog
database is the most obvious choice. Here we have two options. One is
the predicate \term{rdf}{Subject, Predicate, Object} using the argument
types described above. The alternative is to map each RDF predicate on a
Prolog predicate \mbox{\arg{Predicate}({\arg{Subject}, \arg{Object}})}.
We have chosen for \index{rdf/3}\predref{rdf}{3} because it supports queries with uninstantiated
predicates better and a single predicate is easier to manage than an
unbounded set of predicates with unknown names.

\subsection{Input and output of RDF documents}
\label{sec:rdfio}

The RDF/XML parser is realised as a Prolog library on top of the XML
parser described in \secref{markupdom}. Similar to the XML parser it has
two interfaces. The predicate \term{load_rdf}{+Src, -Triples, +Options}
parses a document and returns a Prolog list of \term{rdf}{S,P,O}
triples. Note that despite harvesting to the database is the typical
use-case scenario, the parser delivers a list of triples for maximal
flexibility. The predicate \term{process_rdf}{+Src, :Action, +Options}
exploits the mixed call-back/convert mode of the XML parser to process
the RDF file one \jargon{description} (record) at a time, calling
\arg{Action} with a list of triples extracted from the description.
\Figref{rdfload} illustrates how this is used by the storage module to
load unbounded files with limited stack usage. Source location as
\bnfmeta{file}:\bnfmeta{line} is passed to the \arg{Src} argument of \index{assert_triples/2}\predref{assert_triples}{2}.

\begin{figure}
\figrule
\begin{verbatim}
load_triples(File, Options) :-
        process_rdf(File, assert_triples, Options).

assert_triples([], _).
assert_triples([rdf(S,P,O)|T], Src) :-
        rdf_assert(S, P, O, Src),
        assert_triples(T, Src).
\end{verbatim}

\noindent
    \caption{Loading triples using \index{process_rdf/3}\predref{process_rdf}{3}}
    \label{fig:rdfload}
\figrule
\end{figure}

In addition to named URIs, RDF resources can be \jargon{blank-nodes}. A
blank-node (short \jargon{bnode}) is an anonymous resource that is
created from an in-lined description. \Figref{bnode} describes the
dimensions of a painting as a compound instance of class \arg{Dimension}
with width and height properties. The \arg{Dimension} instance has no
URI. Our parser generates an identifier that starts with a double
underscore, followed by the source and a number. The double underscore
is used to identify bnodes. Source and number are needed to guarantee
the bnode is unique.

\begin{figure}
\figrule
\begin{verbatim}
<Painting rdf:about="...">
  <dimension>
     <Dimension width="45" height="50"/>
  </dimension>
</Painting>

\end{verbatim}

\noindent
    \caption{Blank node to express the compound dimension property}
    \label{fig:bnode}
\figrule
\end{figure}

The parser from XML to RDF triples covers the full RDF specification,
including Unicode handling, RDF datatypes and RDF language tags. The
Prolog source is 1,788 lines. It processes approximately 9,000 triples
per second on an AMD 1600+ based computer. Implementation details and
evaluation of the parser are described in \cite{Wielemaker:03a}.%
	\footnote{The parser described there did not yet support RDF
		  datatypes and language tags, nor Unicode.}

We have two libraries for writing RDF/XML. One,
\term{rdf_write_xml}{+Stream, +Triples}, provides the inverse of
\index{load_rdf/2}\predref{load_rdf}{2}, writing an XML document from a list of \term{rdf}{S,P,O}
terms. The other, called \index{rdf_save/2}\predref{rdf_save}{2} is part of the RDF storage module
described in \secref{rdfdb} and writes a database directly to a file or
stream. The first (\index{rdf_write_xml/2}\predref{rdf_write_xml}{2}) is used to exchange computed graphs
to external programs using network communication, while the second
(\index{rdf_save/2}\predref{rdf_save}{2}) is used to save modified graphs back to file. The resulting
code duplication is unfortunate, but unavoidable. Creating a temporary
graph in a database requires potentially much memory, and harms
concurrency, while graphs fetched from the database into a list may not
fit in the Prolog stacks and is also considerably slower than a direct
write.

\subsection{Storage of RDF}
\label{sec:rdfdb}


Assuming the `harvesting' use-case, we need to implement a
predicate \term{rdf}{?S,?P,?O}. Indexing the database is crucial for
good performance. \Tabref{rdfindex} illustrates the calling pattern from
a real-world application counting 4 million triples. Also note that our data
is described by \figref{rdftypes}. The RDF store was
developed in the context of projects which formulated the following
requirements.

\begin{itemize}
    \item Upto at least 10 million triples on 32-bit hardware.
    \item Fast graph traversal using any instantiation pattern.
    \item Case-insensitive search on literals.
    \item Prefix search on literals for completion in the User Interface.
    \item Searching for words that appear in literals.
    \item Multi-threaded access based on read/write locks.
    \item Transaction management and persistent store.
    \item Maintain source information, so we can update, save or
          remove data based on its source.
    \item Fast load/save of current state.
\end{itemize}

\begin{table}
    \caption{Call-statistics on a real-world system}
    \label{tab:rdfindex}
\begin{tabular}{cccr}
\hline\hline
\multicolumn{3}{c}{\bf Index pattern} & \bf Calls \\
\hline
-   &	-   &	-   &	58 \\
+   &	-   &	-   &	253,554 \\
-   &	+   &	-   &	62 \\
+   &	+   &	-   &	23,292,353 \\
-   &	-   &	+   &	633,733 \\
-   &	+   &	+   &	7,807,846 \\
+   &	+   &	+   &	26,969,003 \\
\hline\hline
\end{tabular}
\end{table}

Our first version of the database used the Prolog database with
secondary tables to improve indexing. As requirements pushed us against
the limits of what is achievable in a 32-bit address-space we decided to
implement the low level store in C. Profiting from the known uniform
structure of the data we realised about two times more compact storage
with better indexing than using a pure Prolog approach. We took the
following design decisions for the C-based storage module:

\begin{itemize}
    \item The RDF \emph{predicates} are represented as unique entities and
          organised according to the rdfs:subPropertyOf relation in
	  multiple hierarchies. The root of each hierarchy is used to
	  compute the hash for the triple. If there is no unique root
	  due to a cycle an arbitrary predicate is assigned to be the
	  root.

    \item Literals are kept in an AVL tree, sorted case-insensitive
	  and case-preserving (e.g. AaBb\ldots).  Numeric literals
	  preceed all non-numeric and are kept sorted on their
	  numeric value.   Storing literals in a separate sorted table
	  allows for indexed search for prefixes and numeric values.  It
	  also allows for monitoring creation and destruction of
	  literals to maintain derived tables such as stemming or
	  double methaphone \cite{Philips:2000:DMS} based on
	  \index{rdf_monitor/3}\predref{rdf_monitor}{3} described below. The space overhead of
	  maintaining the table is roughly cancelled by avoiding
	  duplicates. Experience on real data ranges between -5\% and
	  +10\%.

    \item Resources are represented by Prolog atom-handles.  The hash
	  is computed from the handle-value.  Note that avoiding the
	  translation between Prolog atom and text avoids both
	  duplication of data and table-lookup. We consider
	  this a crucial aspect.

    \item Each triple is represented by the atom-handle for the subject,
	  predicate-pointer, atom-handle or literal pointer for object,
	  a pointer to the source, a line number,
	  a general bit-flag field and 6 `hash-next' pointers covering
	  all indexing patterns except for +,+,+ and +,-,+.  Queries
	  using the pattern +,-,+ are rare.  Fully instantiated
	  queries internally use the pattern +,+,-, assuming few
	  values on the same property.  Considering experience with
	  real data we will probably add a +,+,+ index in the future.
	  The un-indexed table is a simple linked list. The others are
	  hash-tables that are automatically resized if they become too
	  populated.
\end{itemize}

The store itself does not allow for writes while there are active reads
in progress. If another thread is reading, the write operation will stall
until all threads have finished reading. If the thread itself has an
open choicepoint a permission error exception is raised. To arrive at
meaningful update semantics we introduced \jargon{transactions}. The
thread starting a transaction obtains a write-lock, initially allowing
readers to proceed. During the transaction all changes are recorded in a
linked list of actions. If the transaction is ready for commit, the
thread denies access to new readers and waits for all readers to vanish
before updating the database. Transactions are realised by
\term{rdf_transaction}{:Goal}. If \arg{Goal} succeeds, its choicepoints
are discarded and the transaction is committed. If \arg{Goal} fails or
raises an exception the transaction is discarded and \index{rdf_transaction/1}\predref{rdf_transaction}{1}
returns failure or exception. Transactions can be nested. Nesting a
transaction places a transaction-mark in the list of actions of the
current transaction. Committing implies removing this mark from the
list. Discarding removes all action cells following the mark as well as
the mark itself.

It is possible to monitor the database using \term{rdf_monitor}{:Goal,
+Events}.  Whenever one of the monitored events happens \arg{Goal} is
called.  Modifying actions inside a transaction are called during the
commit. Modifications by the monitors are collected in a new transaction
which is committed immediately after completing the preceeding commit.
Monitor events are assert, retract, update, new_literal, old_literal,
transaction begin/end and file-load.  \arg{Goal} is called in the
modifying thread.  As this thread is holding the database write lock,
all invocations of monitor calls are fully serialized.

Although the 9,000 triples per second of the RDF/XML parser ranks it
among the fast parsers, loading 10 million triples takes nearly 20
minutes. For this reason we developed a binary format. The format is
described in \cite{Wielemaker:03a} and loads approximately 10 times
faster than RDF/XML, while using about the same space. The format is
independent from byte-order and word-length, supporting both 32- and
64-bit hardware.

Persistency is achieved through the library \file{rdf_persistency.pl},
which uses \index{rdf_monitor/3}\predref{rdf_monitor}{3} to maintain a set of files in a directory. Each
source known to the database is represented by two files, one file
representing the initial state using the quick-load binary format and
one file containing Prolog terms representing changes, called the
\jargon{journal}.

\subsection{Reasoning with RDF documents}
\label{sec:rdfreasoning}

We have identified two approaches for reasoning on top of the plain RDF
predicate for more high-level languages such as RDFS or OWL. One
approach is taken by the SeRQL query system described in \secref{serql}.
It is based on the observation that these languages provide rules to
deduce new triples from the set of known triples. The API for high level
languages is now simply the \index{rdf/3}\predref{rdf}{3} predicate, where \term{rdf}{S,P,O} is
true for any triple in the deductive closure of the original triple set
under the given language. The deductive closure can be realised using
full forward reasoning, deducing new triples until this is no longer
possible or by a combination of backward reasoning and forward
reasoning. An alternative approach is to consider RDFS or OWL at the
conceptual level and introduce a set of predicates that are inspired on
this level. This approach is taken by our library \file{rdfs.pl},
defining predicates such as \term{rdfs_individual_of}{?Resource,
?Class}, \term{rdfs_subclass_of}{?Sub, ?Super}. \Figref{rdfentail}
illustrates the difference in these approaches.

\begin{figure}
\figrule
\hfill
\begin{minipage}[t]{.32\textwidth}
\begin{verbatim}
% triples
mary type woman .
woman type Class .
woman subClassOf human .
human type Class .
\end{verbatim}

\noindent
\end{minipage}
\hfill
\begin{minipage}[t]{.30\textwidth}
\begin{verbatim}
% entailment interface
?- rdf(mary, type, X).
X = woman ;
X = human ;
No
\end{verbatim}

\noindent
\end{minipage}
\hfill
\begin{minipage}[t]{.35\textwidth}
\begin{verbatim}
% RDFS interface
?- rdfs_individual_of(mary, X).
X = woman ;
X = human ;
No
\end{verbatim}

\noindent
\end{minipage}
\hfill
    \caption{Different interface styles for RDFS}
    \label{fig:rdfentail}
\figrule
\end{figure}

\subsection{Experience}

The RDF infrastructure is used in two of the three case-studies
described at the end of this paper. RDF has a natural representation in
Prolog, either as a list of terms or as a pure predicate. Prolog
non-determinism greatly simplifies querying the database. Transitivity
is easily expressed using recursion. However, as cycles in SW graphs are
allowed and frequent, such algorithms must be protected against
them. Cycle-detection complicates the code and harms performance. We
plan to investigate tabling \cite{iclp95*697} to improve on this
situation.

Although designed as RDF store for SW-based projects, the infrastructure
is also commonly used to create RDF documents from other sources as well
as for filtering and reorganizing RDF documents. In the e-culture
project\fnurl{e-culture.multimedian.nl} it has been used to convert
WordNet \cite{Miller:95a} from its Prolog representation and the Getty
thesauri\fnurl{http://www.getty.edu/research/conducting_research/vocabularies/}
from XML (4GB data) into RDF.



\section{Supporting HTTP}
\label{sec:http}

HTTP, or HyperText Transfer Protocol, is the key W3C standard protocol
for exchanging Web documents. All browsers and Web servers implement it.
The initial version of the protocol was very simple. The client request
consists of a single line of the format \mbox{\bnfmeta{action} \bnfmeta{path}}, the
server replies with the requested document and closes the connection.
Version 1.1 of the protocol is more complicated, providing additional
name-value pairs in the request as well as the reply, features to
request status such as modification time, transfer partial documents,
etcetera.

Adding HTTP support in Prolog, we must consider both the client- and
server-side. In both cases our choice is between doing it in Prolog or
re-using an existing application or library by providing an interface for
it. We compare our work with PiLLoW \cite{journals/corr/cs-DC-0312031} and
the ECLiPSe HTTP services \cite{Eclipse-http}.

Given a basic TCP/IP socket library, writing an HTTP client is trivial
(our client is 258 lines of code). Both PiLLoW and ECLiPSe include
a client written in Prolog.   More issues complicate the choice for a
pure Prolog based server.  

\begin{itemize}
    \item
The server is more complex, which implies there is more to gain by
re-using external code.  Our core server library counts 1,784 lines.

    \item
A single computer can only host one server at port 80 used by default
for public HTTP. Using an alternate port for middleware and storage tier
components is no problem, but use as a public server often conflicts
with firewall or proxy settings. This can be solved using a proxy
server such as the Apache
\emph{mod_proxy}\fnurl{http://httpd.apache.org/docs/1.3/mod/mod_proxy.html}
configured as \jargon{reverse proxy}.

    \item
Servers by definition introduce security risks.  Administrators are
reluctant to see non-proven software in the role of a public server.
Using a proxy as above also reduces this risk, especially if the 
proxy blocks malformed requests.
\end{itemize}

Despite these observations, we consider, like the ECLiPSe team, a pure
Prolog based server worthwhile. As argued in \secref{mt}, many Prolog
Web applications profit from using state stored in the server. Large
resources such as WordNet \cite{Miller:95a} cause long startup times. In
such cases the use of CGI (Common Gateway Interface) is not appropriate
as a new copy of the application is started for each request. PiLLoW
resolves this issue by using \jargon{Active Modules}, where a small CGI
application talks to a continuously running Prolog server using a
private protocol. Using a Prolog HTTP server and optionally a reverse
proxy has the same benefits, but based on a standard protocol, it is
much more flexible.

Another approach is embedding Prolog in another server framework such as
the Java based Tomcat server. Although feasible, embedding non-Java
based Prolog systems in Java is complicated. Embedding through
\emph{jni} introduces platform and Java version dependent problems.
Connecting Prolog and Java concurrency models and garbage collection is
difficult and the resulting system is much harder to manage by the user
than a pure Prolog based application.

\begin{figure}
    \centerline{\includegraphics[width=0.7\linewidth]{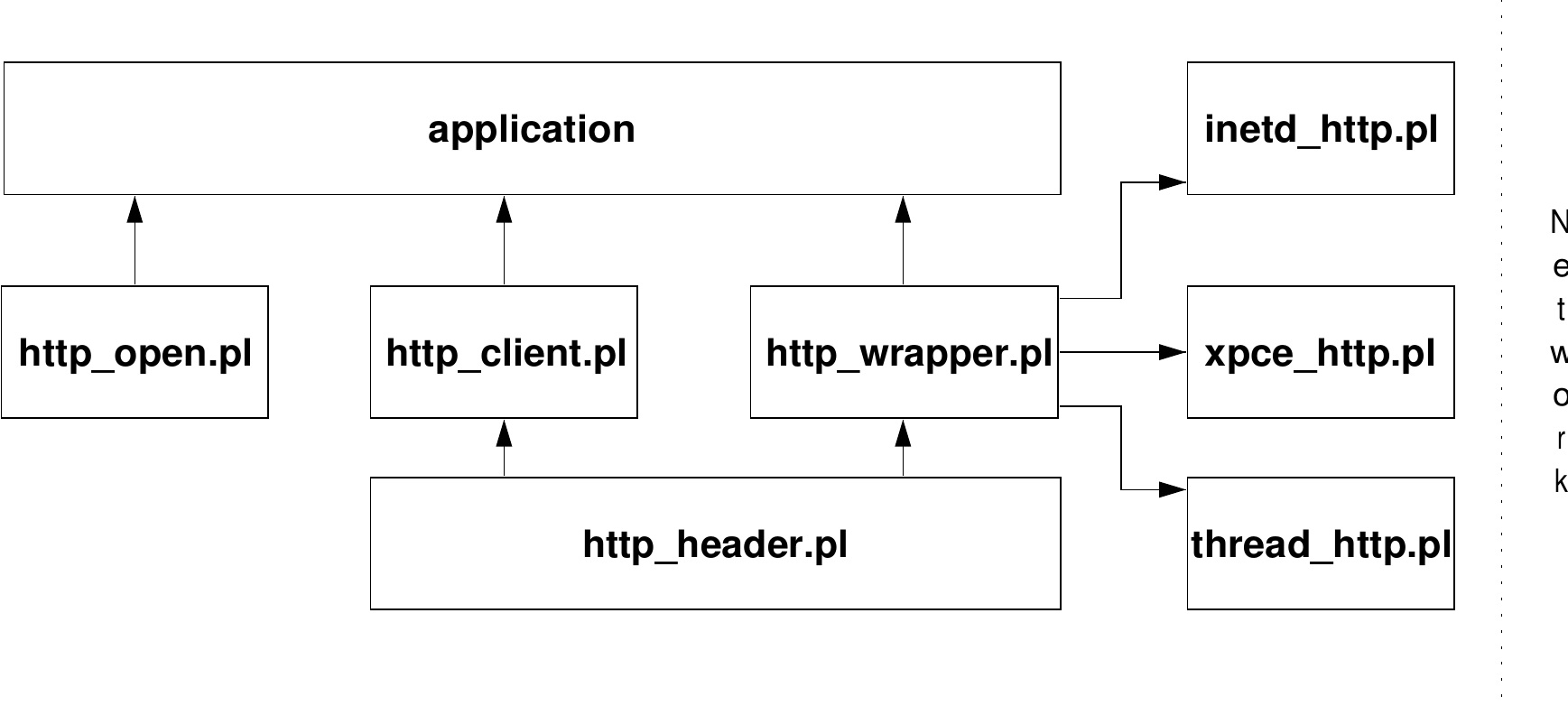}}
    \caption{Module dependencies of the HTTP library}
    \label{fig:http}
\end{figure}

In the following sections we describe our HTTP client and server
libraries.  An overall overview of the modules and their dependencies is
given in \figref{http}.

\subsection{HTTP client libraries}
\label{sec:httpclient}

We support two clients. The first is a lightweight client that only
supports the HTTP GET method by means of \term{http_open}{+URL, -Stream,
+Options}. \arg{Options} allows for setting a timeout or proxy as well
as getting information from the reply-header such as the size of the
document. The \index{http_open/3}\predref{http_open}{3} predicate internally handles HTTP 3xx
(redirect) replies. Other non-ok replies are mapped to a Prolog
exception. After reading the document the user must close the returned
stream-handle using the standard Prolog \index{close/1}\predref{close}{1} predicate. This
predicate makes accessing an HTTP resource as simple as accessing a
local file. The second library, called \file{http_client.pl}, provides
support for HTTP POST and a plugin interface that allows for installing
handlers for documents of specified MIME-types. It shares
\file{http_header.pl} with the server libraries for DCG based creation
and parsing of HTTP headers. Currently provided plugins include
\file{http_mime_plugin.pl} to handle multipart MIME messages and
\file{http_sgml_plugin.pl} for automatically parsing HTML, XML and SGML
documents. \Figref{httpclient} shows the code for fetching a URL and
parsing the returned HTML document it into a Prolog term as described in
\secref{markupdom}.

\begin{figure}
\figrule
\begin{verbatim}
?- use_module(library('http/http_client')).
?- use_module(library('http/http_sgml_plugin')).

?- http_get('http://www.swi-prolog.org/', DOM, []).
DOM = [element(html, [version='-//W3C//DTD HTML 4.0 Transitional//EN'],
                     [element(head, [],
                              [element(title, [],
                                       ['SWI-Prolog\'s Home']), ...
\end{verbatim}

\noindent
    \caption{Fetching an HTML document}
    \label{fig:httpclient}
\figrule
\end{figure}

Both the PiLLoW and ECLiPSe approach return the documents content as a
string. Our interface is stream-based (\index{http_open/3}\predref{http_open}{3}) or allows for
plugin-based processing of the stream (\index{http_get/3}\predref{http_get}{3}, \index{http_post/4}\predref{http_post}{4}). This
interface avoids potentially large intermediate data-structures and
allows for processing unbounded documents.

\subsection{The HTTP server library}
\label{sec:httpd}

\begin{figure}
\hfill
\begin{minipage}[c]{.45\textwidth}
\footnotesize
\begin{verbatim}
:- use_module(
   library('http/thread_httpd')).

start_server(Port) :-
  http_server(reply, [port(Port)]).

reply(Request) :-
  format('Content-type: text/plain~n~n'),
  writeln(Request).
\end{verbatim}

\noindent
\end{minipage}
\hfill
\begin{minipage}[c]{.50\textwidth}
\includegraphics[width=\linewidth]{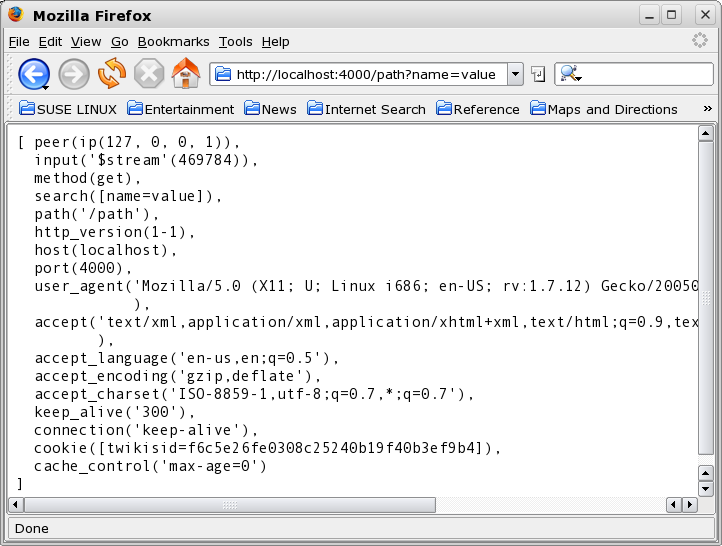}
\end{minipage}
\hfill

    \caption{A simple HTTP server.  The right window shows the client
	     and the format of the parsed request.}
    \label{fig:httpreply}
\end{figure}

Both to simplify re-use of application code and to make it possible to
use the server without committing to a large infrastructure we adopted
the reply-strategy of the CGI protocol, where the handler writes a page
consisting of an HTTP header followed by the document content.
\Figref{httpreply} provides a simple example that returns the
request-data to the client. By importing \file{thread_http.pl} we
implicitly selected the multi-threaded server model. Other models
provided are \file{inetd_http}, causing the (Unix) inet daemon to start
a server for each request and \file{xpce_http} which uses I/O
multiplexing realising multiple clients without using Prolog threads.
The logic of handling a single HTTP request given a predicate realising
the handler, an input and output stream is implemented by
\file{http_wrapper}.

Replies other than ``200 OK'' are generated using a Prolog exception.
Recognised replies are defined by the predicate
\term{http_reply}{+Reply, +Stream, +Header}. For example to indicate that the
user has no access to a page we must use the following call. 

\begin{verbatim}
        throw(http_reply(forbidden(URL))).
\end{verbatim}

\noindent
Failure of the handler raises a ``404 existence error'' reply, while
exceptions other than the ones described above raise a ``500 Server
error'' reply.

\subsubsection{Form parameters}
\label{sec:httpparam}

The library \file{http_parameters.pl} defines
\term{http_parameters}{+Request, ?Parameters} to fetch and type-check
parameters transparently for both GET and POST requests.
\Figref{httpparms} illustrates the functionality. Parameter values are
returned as atoms. If large documents are transferred using a POST
request the user may wish to revert to \term{http_read_data}{+Request,
-Data, +Options} underlying \index{http_get/3}\predref{http_get}{3} to process arguments using
plugins.

\begin{figure}
\figrule
\begin{verbatim}
reply(Request) :-
        http_parameters(Request,
                        [ title(Title, [optional(true)]),
                          name(Name,   [length >= 2]),
                          age(Age,     [integer])
                        ]), ...
\end{verbatim}

\noindent
    \caption{Fetching HTTP form data}
    \label{fig:httpparms}
\figrule
\end{figure}

\subsubsection{Session management}
\label{sec:httpsession}

The library \file{http_session.pl} provides session over the stateless
HTTP protocol. It does so by adding a cookie using a randomly generated
code if no valid session id is found in the current request. The
interface to the user consists of a predicate to set options (timeout,
cookie-name and path) and a set of wrappers around \index{assert/1}\predref{assert}{1} and
\index{retract/1}\predref{retract}{1}, the most important of which are
\term{http_session_assert}{+Data}, \term{http_session_retract}{?Data}
and \term{http_session_data}{?Data}.  In the current version the data
associated with sessions that have timed out is simply discarded.  
Session-data does not survive the server.

Note that a session generally consists of a number of HTTP requests and
replies. Each \emph{request} is scheduled over the available worker
threads and requests belonging to the same session are therefore
normally not handled by the same thread. This implies no session state
can be stored in global variables or in the control-structure of a
thread. If such style of programming is wanted the user must create a
thread that represents the session and setup communication from the
HTTP-worker thread to the session thread. \Figref{httpsession}
illustrates the idea.

\begin{figure}
\figrule
\begin{verbatim}
reply(Request) :-                               % HTTP worker
        (   http_session_data(thread(Thread))
        ->  true
        ;   thread_create(session_loop([]), Thread, [detached(true)]),
            http_session_assert(thread(Thread))
        ),
        current_output(CGIOut),
        thread_self(Me),
        thread_send_message(Thread, handle(Request, Me, CGIOut)),
        thread_get_message(_Done).

session_loop(State) :-                          % Session thread
        thread_get_message(handle(Request, Sender, CGIOut)),
        next_state(Request, State, NewState, CGIOut).
        thread_send_message(Sender, done).
\end{verbatim}

\noindent
    \caption{Managing a session in a thread. The \index{reply/1}\predref{reply}{1} predicate is
	     part of the HTTP worker pool, while \index{session_loop/1}\predref{session_loop}{1} is
	     executed in the thread handling the session.  We omitted
	     error handling for readability of the example.}
    \label{fig:httpsession} \figrule
\end{figure}

\subsubsection{Evaluation}
\label{sec:httpevaluation}

The presented server infrastructure is currently used by many internal
and external projects. Coding a server is very similar to writing CGI
handlers and running in the interactive Prolog process is much easier
to debug. As Prolog is capable of reloading source files in the running
system, handlers can be updated while the server is running. Handlers
running during the update are likely to die on an exception though. We
plan to resolve this issue by introducing read/write locks. The protocol
overhead of the multi-threaded server is illustrated in
\tabref{httpperf}.

\begin{table}
    \caption{HTTP performance executing a trivial query 10,000 times.
	     Times are in seconds. Localhost, dual AMD 1600+ running
	     SuSE Linux 10.0}
    \label{tab:httpperf}
\begin{tabular}{lrrr}
\hline\hline
\bf Connection & \bf Elapsed & \bf Server CPU & \bf Client CPU \\
\hline
Close 	       & 20.84	    & 11.70	     & 7.48 \\
Keep-Alive     & 16.23      & 8.69           & 6.73 \\
\hline\hline
\end{tabular}
\end{table}


\section{Enabling extensions to the Prolog language}
\label{sec:prologext}

SWI-Prolog has been developed in the context of projects many of which
caused the development to focus on managing Web documents and protocols.
In the previous sections we have described our Web enabling libraries.
In this section we describe extensions to the ISO-Prolog standard
\cite{Deransart:96} we consider crucial for scalable and comfortable
deployment of Prolog as an agent in a Web centred world.

\subsection{Multi-threading}
\label{sec:mt}


Concurrency is necessary for applications for the following reasons:

\begin{itemize}
    \item
Network delays may cause communication of a single transaction to take
very long. It is not acceptable if such incidents block access for other
clients. This can be achieved using multiplexed I/O, multiple processes
handling requests in a pool or multiple threads in one or more processes
handling requests in a pool.

    \item 
CPU intensive services must be able to deploy multiple CPUs. This can be
achieved using multiple instances of the service and load-balancing or a
single server running on multi-processor hardware or a combination of
the two.
\end{itemize}

As indicated, none of the requirements above require multi-threading
support in Prolog. Nevertheless, we added multi-threading
\cite{Wielemaker:03c} because it resolves the problems mentioned above
for medium-scale applications while greatly simplifying deployment and
debugging in a platform independent way. A multi-threaded server also
allows maintaining state for a specific session or even shared between
multiple sessions simply in the Prolog database. The advantages of this
are described in \cite{aurora-www}, using the or-parallel Aurora to
serve multiple clients. This is particularly interesting for accessing
the RDF database described in \secref{rdfdb}.

\subsection{Atoms and Unicode support}
\label{sec:unicode}

\begin{figure}
    \centerline{\includegraphics[width=0.3\linewidth]{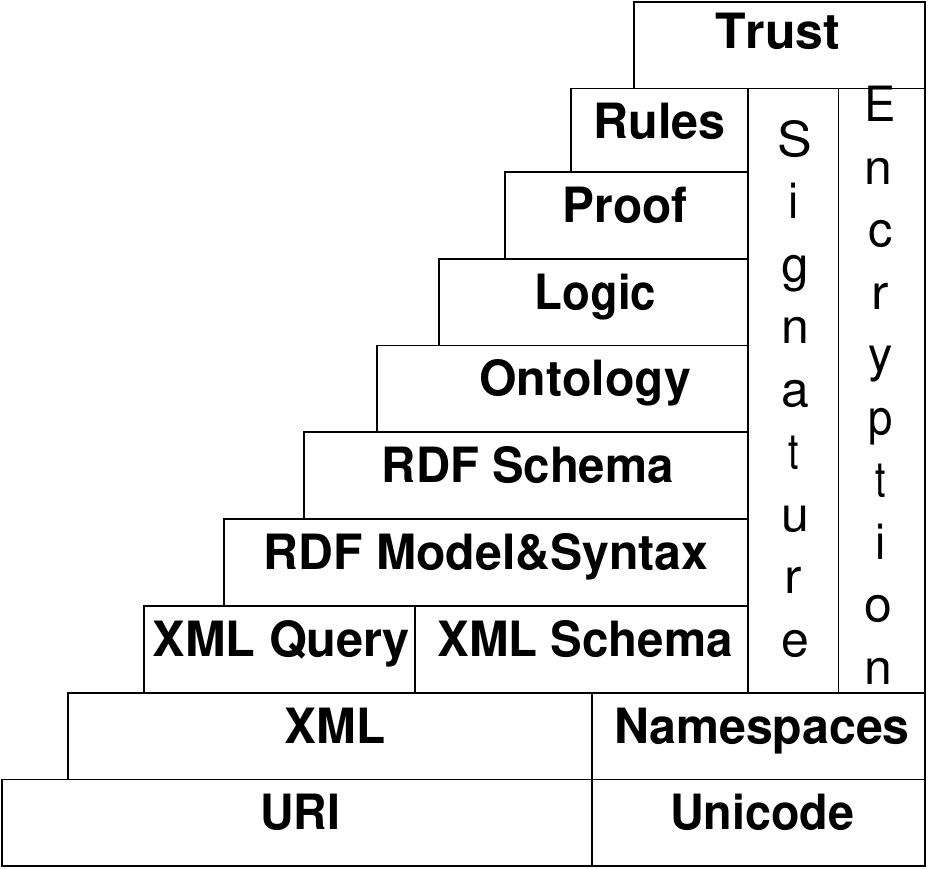}}
    \caption{The Semantic Web layer cake by Tim Burners Lee}
    \label{fig:swcake}
\end{figure}

Unicode\footnote{\url{http://www.Unicode.org/}} is a character encoding
system that assigns unique integers (code-points) to all characters of
almost all scripts known in the world. In Unicode 4.0, the code-points
range from 1 to 0x10FFFF. Unicode can handle documents from different
scripts as well as documents that contain multiple scripts in a single
uniform representation, an important feature in applications processing
Web data. Traditional HTML applications commonly insert special symbols
through entities such as the copyright (\copyright) sign, Greek and
mathematical symbols, etcetera. Using Unicode we can represent all
entity values as plain text. As illustrated in the famous
\jargon{Semantic Web layer cake} in \figref{swcake}, Unicode is at the
heart of the Semantic Web.

\urldef{\unicodeuri}\url{http://www.w3.org/TR/rdf-concepts/#section-Graph-URIref}%
HTML documents can be represented using Prolog strings because Prolog
integers can represent all Unicode code-points. As we have claimed in
\secref{markupdom} however, using Prolog strings is not the most obvious
choice. XML attribute names and values can contain arbitrary Unicode
characters, which requires the unnatural use of strings for these as
well. If we consider RDF, URIs can have arbitrary Unicode
characters\footnote{\unicodeuri} and we want to represent URIs as atoms
to exploit compact storage as well as fast equivalence testing. Without
Unicode support in atoms we would have to encode Unicode in the atom
using escape sequences. All this patchwork can be avoided if we demand
the properties below for Prolog atoms.

\begin{itemize}
    \item Atoms represent text in Unicode
    \item Atoms have no limit on their length
    \item The Prolog implementation allows for a large number of
          atoms, both to represent URIs and to represent text in
	  HTML/XML documents. SWI-Prolog's limit is $2^{25}$ 
	  (32 million).
    \item Continuously running servers cannot allow memory leaks
          and therefore processing dynamic data using atoms
	  requires atom garbage collection.
\end{itemize}


\section{Case study --- A Semantic Web Query Language}
\label{sec:serql}

\begin{figure}
    \centerline{\includegraphics[width=0.7\linewidth]{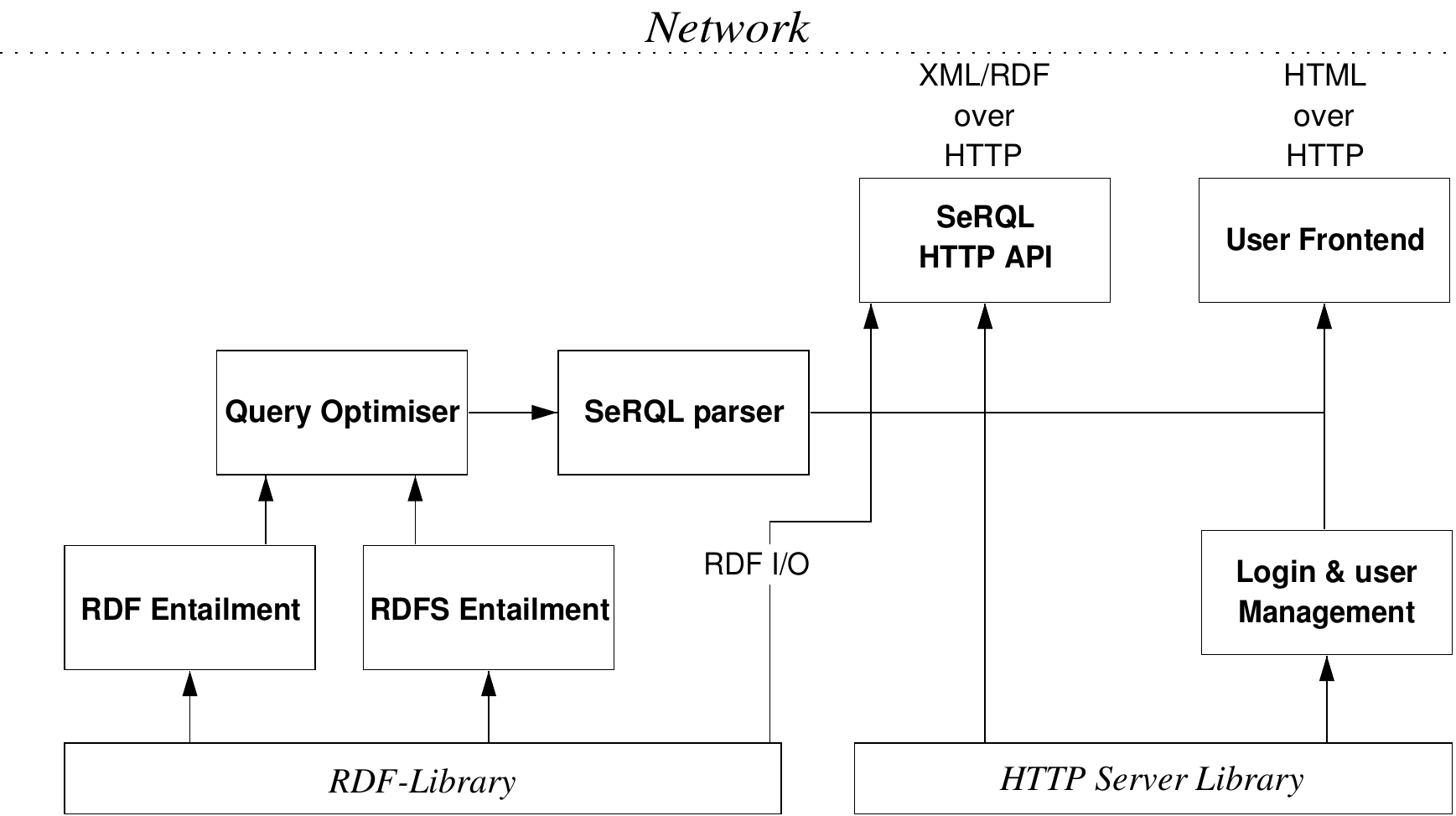}}
    \caption{Module dependencies of the SeRQL system.  Arrows denote
	     `imports from' relations.}
    \label{fig:serql}
\end{figure}

In this case-study we describe the SWI-Prolog SeRQL implementation.%
    \footnote{\url{http://www.swi-prolog.org/packages/SeRQL}}
SeRQL is an RDF query language developed as part of the Sesame project%
    \footnote{\url{http://www.openrdf.org}} \cite{Broekstra:02a}.
SeRQL uses HTTP as its access protocol. Sesame consists of an
implementation of the server as a Java servlet and a Java
client-library. By implementing a compatible framework we made our
Prolog based RDF storage and reasoning engine available to Java clients.
The Prolog SeRQL implementation uses all of the described SWI-Prolog
infrastructure and building it has contributed significantly to the
development of the infrastructure. \Figref{serql} lists the main
components of the server.

The \jargon{entailment} modules are plugins that implement the
entailment approach to RDF reasoning described in \secref{rdfreasoning}.
They implement \index{rdf/3}\predref{rdf}{3} as a pure predicate, adding implicit triples to the
raw triples loaded from RDF/XML documents. \Figref{rdfentail2} shows the
somewhat simplified entailment module for RDF. The multifile rule
registers the module as entailment module for the SeRQL system. New
modules can be loaded dynamically into the platform, providing support
for other SW languages or application-specific server-side reasoning.
Prolog's dynamic loading and re-loading allows for updating such
reasoning modules on the live server.

\begin{figure}
\figrule
\begin{verbatim}
:- module(rdf_entailment, [rdf/3]).

rdf(S, P, O) :-
        rdf_db:rdf(S, P, O).
rdf(S, rdf:type, rdf:'Property') :-
        rdf_db:rdf(_, S, _),
        \+ rdf_db:rdf(S, rdf:type, rdf:'Property').
rdf(S, rdf:type, rdfs:'Resource') :-
        rdf_db:rdf_subject(S),
        \+ rdf_db:rdf(S, rdf:type, rdfs:'Resource').

:- multifile serql:entailment/2.
serql:entailment(rdf, rdf_entailment).
\end{verbatim}

\noindent
    \caption{RDF entailment module}
    \label{fig:rdfentail2}
\figrule
\end{figure}

The SeRQL parser is a DCG-based parser translating a SeRQL query
text into a compound goal calling \index{rdf/3}\predref{rdf}{3} and predicates from the
SeRQL runtime library which provide comparison and functions built into
the SeRQL language. The resulting control-structure is passed to the
query optimiser \cite{Wielemaker:05a} which uses statistics maintained
by the RDF database to reorder the pure \index{rdf/3}\predref{rdf}{3} calls for best
performance. The optimiser uses a generate-and-evaluate approach to find
the optimal order. Considering the frequently long conjunctions of \index{rdf/3}\predref{rdf}{3}
calls, the conjunction is split into independent parts.
\Figref{rdfsplit} illustrates this in a very simple example. During
abstract execution, information on instantiation and types implied by
the runtime library predicates is attached to the variables using
dynamic attributed variables. \cite{Demoen:CW350}. 

\begin{figure}
\figrule
\begin{verbatim}
        ...
        rdf(Paper, author, Author),
        rdf(Author, name, Name),
        rdf(Author, affiliation, Affil),
        ...
\end{verbatim}

\noindent
    \caption{Split rdf conjunctions.  After executing the first \index{rdf/3}\predref{rdf}{3}
	     query \arg{Author} is bound and the two subsequent queries
	     become independent.  This is also true for other orderings,
	     so we only need to evaluate 3 alternatives instead of 3!
	     (6).}
    \label{fig:rdfsplit}
\figrule
\end{figure}

HTTP access consists of two parts. The human-centred portal consists of
HTML pages with forms to administer the server as well as view
statistics, load and unload documents and run SeRQL queries
interactively presenting the result as an HTML table. Dynamic pages are
generated using the \file{html_write.pl} library described in
\secref{htmlwrite}. Static pages are served from HTML files by the
Prolog server. Machines use HTTP POST requests to provide query data and
get a reply in XML or RDF/XML.

The system knows about various RDF input and output formats. To reach
modularity the kernel exchanges RDF graphs as lists of terms
\term{rdf}{S,P,O} and result-tables as lists of terms using the functor
\const{row} and arity equal to the number of columns in the table. The
system calls a multifile predicate using the format identifier and data
to realise the requested format. The HTML output format uses
\file{html_write.pl}. The RDF/XML format uses \index{rdf_write_xml/2}\predref{rdf_write_xml}{2} described
in \secref{rdfio}. Both \index{rdf_write_xml/2}\predref{rdf_write_xml}{2} and the other XML output format
use straight calls \index{format/3}\predref{format}{3} to write the document, where quoting values
is realised by quoting primitives provided by the SGML/XML parser
described in \secref{markupdom}. Using direct writing instead of
techniques described in \secref{htmlwrite} avoids potentially large
intermediate datastructures and is not very complicated given the very
simple structure of the documents.

\subsection{Evaluation}

The SeRQL server and the SWI-Prolog library development is too closely
integrated to use it as an evaluation of the functionality provided
by the Web enabling libraries. We compared our server to Sesame, written
in Java. The source code of the Prolog based server is 6,700 lines,
compared to 86,000 for Sesame. As both systems have very different
coverage in functionality and can re-use libraries at different
levels it is hard to judge these figures. Both answer trivial queries in
approximately 5ms on a dual AMD 1600+ PC running Linux 2.6. On complex
queries the two systems perform very differently. Sesame's forward
reasoning makes it handle some RDFS queries much faster. Sesame does not
contain a query optimizer which cause order-dependent and sometimes very
long response times on large conjunctions.

The power of LP where programs can be handled as data is exploited by
parsing the SeRQL query into a program, optimizing the program by
manipulating it as data, after which we can simply call it to answer the
query. The non-deterministic nature of \index{rdf/3}\predref{rdf}{3} allows for a trivial
translation of the query to a non-deterministic program that produces
the answers on backtracking.

The server only depends on the standard SWI-Prolog distribution and
therefore runs unmodified on all systems supporting SWI-Prolog. It has
been tested on Windows, Linux and MacOS X.

All infrastructure described is used in the server. We use \index{format/3}\predref{format}{3},
exploiting XML quoting primitives provided by the Prolog XML library to
print highly repetitive XML files such as the SeRQL result-table.
Alternatively we could have created the corresponding DOM term and
call \index{xml_write/2}\predref{xml_write}{2} from the library \file{sgml_write.pl}.


\section{Case study --- XDIG}
\label{sec:xdig}


In \secref{serql} we have discussed the case study how SWI-Prolog is
used for a RDF query system, i.e., a meta-data management and reasoning
system. In this section we describe a Prolog-powered system for ontology
management and reasoning based on Description Logics (DL). DL has
greatly influenced the design of the W3C ontology language OWL. The DL
community, called DIG (DL Implementation Group) have developed a
standard for accessing DL reasoning engines called the DIG description
logic interface\fnurl{http://dl.kr.org/dig/} \cite{dig}, DIG interface
for short. Many DL reasoners like Racer \cite{racer} and FACT
\cite{fact} support the DIG interface, allowing for the construction of
highly portable and reusable DL components or extensions.

In this case study, we describe XDIG, an eXtended DIG Description Logic
interface, which has been implemented on top of the SWI-Prolog Web
libraries. The DIG interface uses an XML-based messaging protocol on top
of HTTP. Clients of a DL reasoner communicate by means of HTTP POST
requests. The body of the request is an XML encoded message which
corresponds to the DL concept language. Where OWL is based on the triple
model described in \secref{rdfdoc}, DIG statements are grounded
directly in XML. \Figref{madcowstatement} shows a DIG statement which
defines the concept \emph{MadCow} as a cow which eats brains, part of
sheep.

\begin{figure}
\figrule
\begin{verbatim}
<equalc>
  <catom name='mad+cow'/>
  <and>
    <catom name='cow'/>
    <some>
      <ratom name='eats'/>
      <and>
        <catom name='brain'/>
        <some>
          <ratom name='part+of'/>
          <catom name='sheep'/>
</some></and></some></and></equalc>
\end{verbatim}
   \caption{a DIG statement on MadCow}
   \label{fig:madcowstatement}
\figrule
\end{figure}

\subsection{Architecture of XDIG}

The XDIG libraries form a framework to build DL reasoners that have
additional reasoning capabilities. XDIG serves as a regular DL reasoner
via its corresponding DIG interface. An intermediate XDIG server can
make systems independent from application specific characteristics. A
highly decoupled infrastructure significantly improves the reusability
and applicability of software components.

\begin{figure}
\centering
\includegraphics[width=0.8\textwidth]{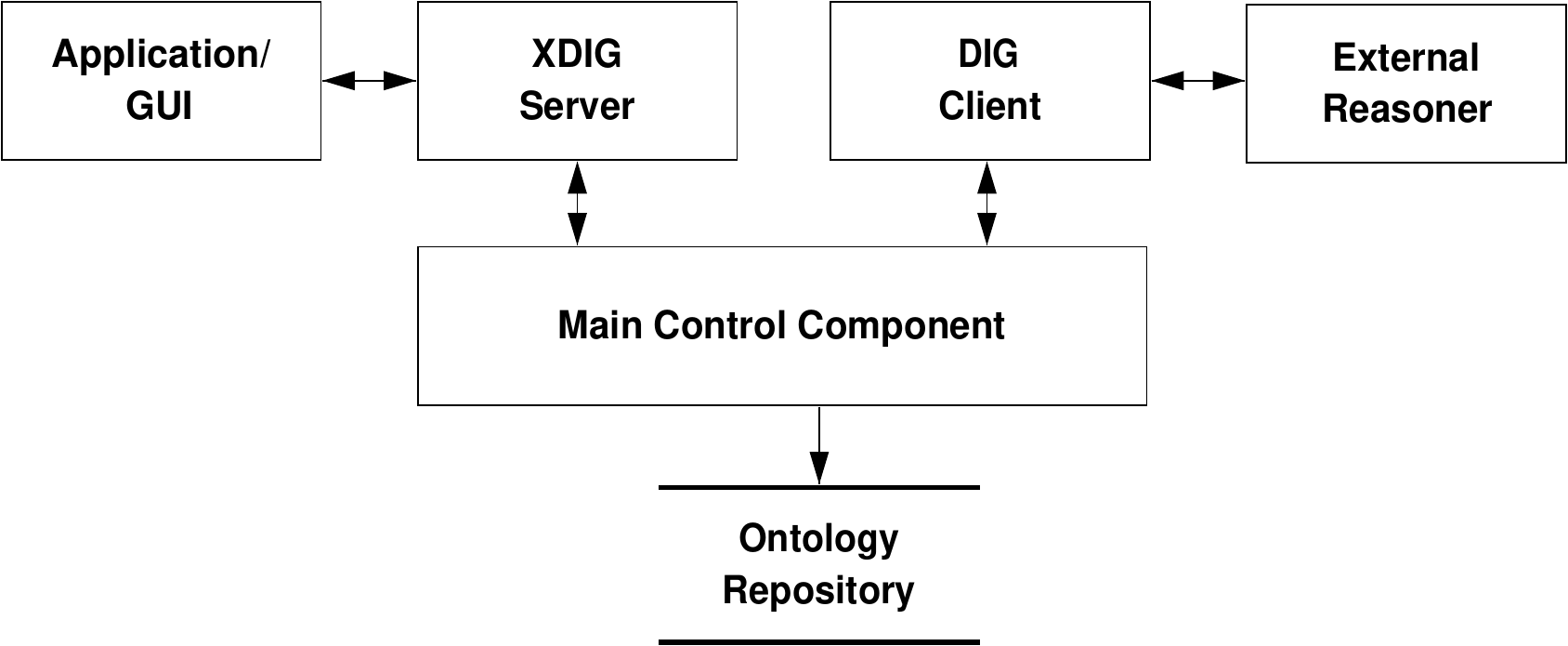}
\caption{Architecture of XDIG}
\label{fig:xdig}
\end{figure}

The general architecture of XDIG is shown in \figref{xdig}. It consists
of the following components:

\begin{description}
  \item [XDIG Server]
The XDIG server deals with requests from ontology applications. It
supports our extended DIG interface, i.e., it not only supports standard
DIG/DL requests, like 'tell' and 'ask', but also additional processing
features like changing system settings. The library \file{dig_server.pl}
implements the XDIG protocol on top of the Prolog HTTP server described
in \secref{httpd}. The predicate \term{dig_server}{+Request} is called
from the HTTP server to process a client's \arg{Request} as illustrated
in \figref{httpreply}. XDIG server developers have to define the
predicate \term{my_dig_server_processing}{+Data, -Answer, +Options},
where \arg{Data} is the parsed DIG XML requests and \arg{Answer} is term
\term{answer}{-Header, -Reply}. \arg{Reply} is the XML-DOM term
representing the answer to the query.

  \item [DIG Client]
XDIG is designed to rely on an external DL reasoner. It implements a
regular DIG interface client and calls the external DL reasoner to
access the standard DL reasoning capabilities. The predicate
\term{dig_post}{+Data, -Reply, +Options} posts the data to the external
DIG server. The predicates are defined in terms of the predicate
\predref{http_post}{4} and others in the HTTP and XML libraries.

\item [Main Control Component]
The library \file{dig_process.pl} provides facilities to analyse DIG
statements such as finding concepts, individuals and roles, but also
decide of \jargon{satisfiability} of concepts and consistency. Some of
this processing is done by analysing the XML-DOM representation of DIG
statements in the local repository, while satisfiability and consistency
checking is achieved by accessing external DL reasoners through the
DIG client module. 

\item [Ontology Repository]
The Ontology Repository serves as an internal knowledge base (KB), which
is used to store multiple ontologies locally. These ontology statements
are used for further processing when the reasoner receives an `ask'
request. The main control component usually selects parts from the
ontologies to post them to an external DL reasoner and obtain the
corresponding answers. This internal KB is also used to store system
settings.
\end{description}

As DIG statements are XML based, XDIG stores statements in the local
repository using the XML-DOM representation described in
\secref{markupdom}. The tree model of XDIG data has been proved to be
convenient for DIG data management.

\Figref{pioncode} shows a piece of code from the library XDIG
defining the predicate \term{direct_concept_relevant}{+DOM, ?Concept}
which checks if a set of \arg{Statements} is directly relevant to a
\arg{Concept}, namely the \arg{Concept} appears in the body of a
statement in the list. The predicate
\predref{direct_concept_relevant}{2} has been used to develop PION for
reasoning with inconsistent ontologies, and DION for inconsistent
ontology debugging.

\begin{figure}
\figrule
\begin{verbatim}
direct_concept_relevant(element(catom, Atts, _), Concept) :-
    memberchk(name=Concept, Atts).
direct_concept_relevant(element(_, _, Content),  Concept) :-
    direct_concept_relevant(Content, Concept).
direct_concept_relevant([H|T], Concept) :-
    (   direct_concept_relevant(H, Concept)
    ;   direct_concept_relevant(T, Concept)
    ).
\end{verbatim}
    \caption{direct_concept_relevant checks that
    a concept is referenced by a DIG statement}
    \label{fig:pioncode}
    \figrule
\end{figure}

\subsection{Application}

XDIG has been used to develop several DL reasoning services. PION is a
reasoning system that deals with inconsistent
ontologies\fnurl{http://wasp.cs.vu.nl/sekt/pion}
\cite{hu2004,huangijcai05}. MORE is a multi-version ontology
reasoner\fnurl{http://wasp.cs.vu.nl/sekt/more} \cite{huang-iswc05}. DION
is a debugger of inconsistent
ontologies\fnurl{http://wasp.cs.vu.nl/sekt/dion}\cite{sekt361}. With the
support of an external DL reasoner like Racer, DION can serve as an
inconsistent ontology debugger using a bottom-up approach.


\newcommand{\component}[1]{\emph{#1}}
\newcommand{\componentlater}[1]{{#1}}
\section{Case study --- Faceted browser on Semantic Web 
database integrating multiple collections}~\label{sec:facetb}

In this case study we describe a pilot for the
STITCH-project\fnurl{http://stitch.cs.vu.nl} whose main aim is
studying and finding solutions for the problem of integrating controlled
vocabularies such as thesauri and classification systems in the
Cultural Heritage domain.
The pilot consists of the integration of two collections -- the Medieval 
Illuminations of the Dutch National Library (Koninklijke 
Bibliotheek) and the Masterpieces 
collection from the Rijksmuseum -- and development of 
a user interface for  browsing the merged collections. One 
requirement within the pilot is to use ``standard 
Semantic Web techniques'' during all stages, so as to be able to evaluate 
their added value. An explicit research goal was to evaluate existing
``ontology mapping'' tools.

The problem could be split into three main tasks:
\begin{itemize} 
\item
Gathering data, i.e.\ records of the collections and controlled 
vocabularies they use, and transforming it into RDF.
\item
Establishing semantic links between the vocabularies using off-the-shelf 
ontology mapping tools. 
\item
Building a prototype User Interface (UI) to access (search and browse) the 
integrated collections and experiment with different ways to access them 
using a Web server.
\end{itemize}
SWI-Prolog has been used in all three tasks; to illustrate the use of 
the SWI-Prolog Web libraries in the pilot, in the next section we focus on 
their application in the prototype UI because it is the largest subsystem 
using these libraries.

\subsection{Multi-Faceted Browser}

Multi-Faceted Browsing is a search and browse paradigm where a collection
is accessed by refining multiple (preferably) structured aspects 
-- called facets -- 
of its elements. For the user interface and user interaction we 
have been influenced by the approach of Flamenco~\cite{hearst02finding}. The 
Multi-Faceted Browser is implemented in SWI-Prolog. All data is stored 
in an RDF database, which can be either an external SeRQL repository or an 
in-memory SWI-Prolog RDF database. The system consists of three components, 
\component{RDF-interaction},
which deals with RDF-database storage and access, 
\component{HTML-code generation}, for the creation of Web pages 
and the \component{Web server} component, 
implementing the HTTP server. They are discussed in the 
following sections.

\subsubsection{RDF-interaction}
\label{sec:rdfinteraction}
We first describe the content of the RDF database before explaining how to 
access it. The RDF database contains:
\begin{itemize}
\item
750 records from the 
Rijksmuseum, and 1000 from the Koninklijke Bibliotheek.
\item
RDF representation of the hierarchically structured facets, 
we use SKOS\footnote{\url{http://www.w3.org/2004/02/skos/}}, a 
model dedicated to the represention of controlled vocabularies.
\item
Mappings between SKOS Concept Schemes used in the different collections. 
\item
Portal-specific information as ``\jargon{Site Configuration Objects}'', 
identified by 
URIs with properties defining what collections are part of 
the setup, what facets are shown, and also values for the constant text in the
Web page presentation and other User Interface configuration properties. 
Multiple such Site Configuration Objects may be defined in a repository.
\end{itemize}

The in-memory RDF store contains, depending on the number of mappings
and structured vocabularies that are stored in the database, about
300,000 RDF triples. The Sesame store contains more triples -- 520,000
-- as its RDFS-entailment implementation implies generation of derived
triples (see \secref{serql}).
  
\paragraph{RDF database access}

Querying the RDF store for more complex results based on URL query
arguments consisted of three steps: 1) building SeRQL queries from URL
query arguments, 2) passing them on to the SeRQL-engine, gathering the
result rows and 3) finally post-processing the output, e.g. counting
elements and sorting them. \Figref{srqlex} shows an example of a
generated SeRQL query. Finding matching records involves finding records
annotated by the facet value or by a value that is a hierarchical
descendant of facet value. We implemented this by interpreting records
as instances of SKOS concepts and using the transitive and reflexive
properties of the rdfs:subClassOf property. This explains for example
\verb-{Rec} rdf:type {<http://www.telin.nl/rdf/topia#Paintings>}- in
\figref{srqlex}.

\begin{figure*}
\begin{verbatim}
SELECT  Rec, RecTitle, RecThumb, CollSpec
FROM {SiteId} rdfs:label {"ARIATOPS-NONE"};
            mfs:collection-spec {CollSpec} mfs:record-type {RT};
            mfs:shorttitle-prop {TitleProp};
            mfs:thumbnail-prop {ThumbProp},
{Rec} rdf:type {RT}; TitleProp {RecTitle};
      ThumbProp {RecThumb},
{Rec} rdf:type {<http://www.telin.nl/rdf/topia#AnimalPieces>},
{Rec} rdf:type {<http://www.telin.nl/rdf/topia#Paintings>} 
USING NAMESPACE skos = <http://www.w3.org/2004/02/skos/core#>,
mfs = <http://www.cs.vu.nl/STITCH/pp/mf-schema#><br>
\end{verbatim}
   \caption{An example of a SeRQL query, which returns details of records
matching two facet values (AnimalPieces and Paintings)}
    \label{fig:srqlex}
\figrule
\end{figure*}

The SeRQL-query interface contains timing and debugging facilities
for single queries; for flexibility it provides access to an external SeRQL 
server\footnote{We used the \file{sesame_client.pl} library that provides an 
interface to external SeRQL servers, packaged with
the SWI-Prolog SeRQL library} for which we used
Sesame\footnote{\url{http://www.openrdf.org}}, but also to the 
in-memory store of the SWI-Prolog SeRQL implementation described in \secref{serql}.


\subsubsection{HTML-code generation}

We used the SWI-Prolog \file{html_write.pl} library described in 
\secref{htmlwrite} for our HTML-code generation. There are three distinct kinds of Web 
pages the multi-faceted browser generates, 
the portal access page, the refinement page and the single collection-item  
page. The DCG approach to generating HTML code made it easy to share HTML-code 
generating procedures such as common headers and HTML code for refinement of 
choices. The \componentlater{HTML-code generation} component contains some 140 DCG 
rules (1200 lines
of Prolog code of which 800 lines are DCG rules), part of which are simple 
list-traversing rules such as the example of \Figref{htmlex}.

\begin{figure*}
\begin{verbatim}
objectstr([],_O, _Cols,_Args) --> [].
objectstr([RowObjects|ObjectsList], Offset, Cols, Args) -->
        { Percentage is 100/Cols },
        html(tr(valign(top),\objectstd(RowObjects, Offset, Percentage, Args))),
        { Offset1 is Offset + Cols },
        objectstr(ObjectsList, Offset1, Cols, Args).

objectstd([], _, _, _) --> [].
objectstd([Url|RowObjects], Index, Percentage, Args) -->
       {   ..
           construct_href_index(..., HRef),
           missing_picture_txt(Url, MP)
       },
       html(td(width(Percentage),a([href(HRef)],img([src(Url),alt(MP)])))),
       { Index1 is Index + 1 },
       objectstd(RowObjects, Index1, Percentage, Args).
\end{verbatim}
   \caption{Part of the html code generation for displaying all the images of a 
query result in an HTML table}
    \label{fig:htmlex}
\figrule
\end{figure*}

\subsubsection{Web Server}

The Web server is implemented using the HTTP server library described 
in \secref{httpd}. The \componentlater{Web server} component itself is very 
small. 
It follows 
the skeleton code described in \Figref{httpreply}.
In our case the \index{reply/1}\predref{reply}{1} predicate 
extracts the URL root and 
parameters from the URL.
The \jargon{Site Configuration Object}, which is introduced in 
\secref{rdfinteraction}, is returned by the \componentlater{RDF-interaction} component 
based on the URL root. It is passed on to 
the \componentlater{HTML-code generation} component which generates Web content
 as 
shown in \index{reply_page/1}\predref{reply_page}{1} in \Figref{affiliation}.

\subsection{Evaluation}

This case study shows that SWI-Prolog is effective for building applications in
the context of the Semantic Web. 
In a single month a fully functional prototype 
portal has been created providing structured access to multiple collections. 
The independence of any external libraries and the full support of
all libraries on different platforms made it easy to develop and install in 
different operating systems. All case study software has been tested to 
install and run transparently both on Linux and on Microsoft Windows.

At the start of the pilot project we briefly evaluated existing environments
for creating multi-faceted browsing portals:
We considered the software available from the Flamenco 
Project~\cite{hearst02finding}
and the OntoViews Semantic Portal Creation Tool~\cite{ontoviews}.
The Flamenco software would require developing a
translation from RDF to the Flamenco data representation. OntoViews 
heavily uses Semantic Web techniques, but the software was unnecessarily 
complex for our pilot, requiring a number of external libraries. This together 
with our need for flexible experiments with various setups made 
us decide to build our own prototype. 

The prototype allowed us to easily experiment with and develop various 
interesting ways of providing users access to integrated heterogeneous 
collections~\cite{stitchpilot1}. The pilot project portal is accessible on line for your 
evaluation 
at \url{http://stitch.cs.vu.nl/demo.html}.


\section{Conclusion}
\label{sec:conclusion}

We have presented an overview of the libraries and Prolog language
extensions we have implemented and which we provide to the LP community
as Open Source resources. As the presented libraries cover very
different functionality we have compared our approach with related work
throughout the document. We have demonstrated that Prolog, equipped with
an HTTP server library, libraries for reading and writing markup
documents, multi-threading, Unicode support, unbounded atoms and atom
garbage collection, becomes a flexible component in a multi-tier server
architecture. Automatic memory management and the absence of pointers
and other dangerous language constructs justify the use of Prolog in
security sensitive environments. The middleware, typically dealing with
the application logic is the most obvious tier for exploiting Prolog. In
our case studies we have seen Prolog active as storage component
(\secref{serql}), middleware (\secref{xdig}) and in the presentation
tier (\secref{facetb}).

Development in the near future is expected to concentrate on Semantic
Web reasoning, such as the translation of SWRL rules to logic programs.
Such translations will benefit from tabling to reach at more predictable
response-times and allow for more declarative programs. We plan to add
more support for OWL reasoning, possibly supporting vital relations for
ontology mapping such as \verb|owl:sameAs| in the low-level store. We
also plan to add PSP/PWP-like (\secref{genml}) page-generation
facilities.

\section*{Acknowledgements}

Recent development and preparing this paper was supported by the
MultimediaN\fnurl{www.multimedian.nl} project funded through the BSIK
programme of the Dutch Government. Older projects that provided a major
contribution to this research include the European projects IMAT and
GRASP and the Dutch project MIA. The XDIG case-study is supported by FP6
Network of Excellence European project SEKT (IST-506826). The faceted
browser case-study is funded by CATCH, a program of the Dutch
Organization NWO. We acknowledge the SWI-Prolog community for feedback
on the presented facilities, in particular Richard O'Keefe for his
extensive feedback on SGML support.

\bibliography{plweb}

\end{document}